\documentclass[runningheads]{llncs}
\newcommand{\seq}[1]{{\overrightarrow{#1}}}
\usepackage{graphicx} 
\usepackage{upquote} 
\usepackage[english]{babel}
\usepackage{booktabs}
\usepackage{placeins}

\usepackage{paralist}

\usepackage{booktabs}   %

\usepackage{csquotes}

\usepackage{textcmds}

\usepackage{mathtools}
\usepackage{hyperref}
\hypersetup{hidelinks,
  colorlinks=true,
  allcolors=black,
  pdfstartview=Fit,
  breaklinks=true}
\usepackage[all]{hypcap}

\usepackage[capitalise,nameinlink]{cleveref}
\crefname{section}{Sec.}{Secs.}
\crefname{figure}{Fig.}{Figs.}
\crefname{listing}{\lstlistingname}{\lstlistingname}
\Crefname{listing}{Listing}{Listings}

\usepackage{xspace}

\usepackage{amsmath}   
\usepackage{amssymb}
 
\usepackage{semantic}  
\usepackage{complexity}   
\newclass{\NEXPTIME}{NEXPTIME}
\newclass{\UF}{UF}  
\usepackage{mathpartir}

\usepackage[%
  square,        %
  comma,         %
  numbers,       %
  sort&compress, %
  sectionbib
]{natbib}

\usepackage{listings}

\usepackage[ampersand]{easylist}

\usepackage{afterpage}

\usepackage{doi}
\lstnewenvironment{haskellCode} 
    {
      \lstset{}%
      \csname lst@SetFirstLabel\endcsname
      }
    {
      \csname lst@SaveFirstLabel\endcsname
      }
\mathligprotect{\haskellCode}
\newcommand{\hinline}[1]{\lstinline[language=Haskell]{#1}}

\newcommand{\V}{\mathcal{V}}
\renewcommand{\H}{\mathcal{H}}

\renewcommand{\L}{\mathcal{L}} 
\renewcommand{\C}{\mathcal{C}}
\renewcommand{\M}{\mathcal{M}}

\newcommand{\Hb}[2]{{\mathcal{H}|[ #1 |]_{#2}}}
\newcommand{\Eb}[1]{{\mathcal{E}|[ #1 |]}}
\newcommand{\Lb}[1]{{\mathcal{L}|[ #1 |]}}
\newcommand{\Lvb}[1]{{{\mathcal{L}}|[ #1 |]}}

\newcommand{\Pb}[1]{{\mathcal{P}|[ #1 |]}} 
\renewcommand{\P}{\mathcal{P}}
\newcommand{\F}{\mathcal{F}} 
\usepackage{bbold}
\newcommand{\B}{\mathbb{B}} 

\newcommand{\withConstr}{\mid}

\newcommand{\setbot}{\bot}
\newcommand{\false}{\mathbf{F}}
\newcommand{\true}{\mathbf{T}}

\newcommand{\match}{\ \mathtt{match}\ }
\newcommand{\with}{\ \mathtt{with }\ }
\newcommand{\llet}{\ \mathtt{let }\ }
\newcommand{\iin}{\ \mathtt{in }\ }
\newcommand{\LANG}{$\lambda_\textsc{Match}$}
\newcommand{\FV}{\mathsf{FV}}

\newcommand{\inDomain}{\mathcal{D}}

\newsavebox{\saveboxedarray}

\newenvironment{boxedarray}[1]         
  {\begin{lrbox}{\saveboxedarray}\begin{math}\begin{array}{#1}}
  {\end{array}\end{math}\end{lrbox}\fbox{\usebox{\saveboxedarray}}}

  \newenvironment{inferbox}[0]
  {\begin{minipage}{\textwidth}\begin{mathpar}}
      {\end{mathpar}\end{minipage}}

\begin{document}
\title{Set Constraints, Pattern Match Analysis, and SMT}
\author{Joseph Eremondi\thanks{This material is based on work supported by the NSERC CGS-D scholarship.  }
}
\authorrunning{J. Eremondi}
\institute{University of British Columbia, Vancouver, Canada \\
\email{jeremond@cs.ubc.ca}}
\maketitle              %
\begin{abstract}
Set constraints provide a highly general way to formulate program analyses.
However, solving arbitrary boolean combinations of set constraints is
$\NEXPTIME$-hard. Moreover, while theoretical algorithms to solve
arbitrary set constraints exist, they are either too complex to realistically implement
or too slow to ever run.

We present a translation that converts a set constraint formula into an SMT problem.
Our technique allows for arbitrary boolean combinations of set constraints,
and leverages the performance of modern SMT solvers.
To show the usefulness of unrestricted set constraints,
 we use them to devise a pattern match analysis
for functional languages,
which ensures that missing cases of pattern matches are always unreachable.
We implement our analysis in the Elm compiler and show that our translation is fast enough
to be used in practical verification.   

\end{abstract}
\section{Introduction}

Set constraints are a powerful tool for expressing a large number of program analyses
in a generic way. Featuring recursive equations and inequations over variables denoting sets of values,
set constraints allow us to model the sets of values an expression could possibly take. 
While they were an active area of research in decades prior,
they have not seen widespread adoption. In their most general form, finding solutions
for a conjunction of set constraints is \NEXPTIME-complete. While efficient solvers
have been developed for restricted versions of the set constraint problem~\cite{10.1007/BFb0055513,10.1007/11547662_16}, 
solvers for unrestricted set constraints are not used in practice.

However, since the development of set constraints, there have been significant advances
in solvers for SAT modulo theories (SMT). Although SMT requires exponential time in theory,
solvers such as Z3~\cite{10.1007/978-3-540-78800-3_24} and CVC4~\cite{BCD+11}
are able to solve a wide range of satisfiability problems in practice.
Given the success of SMT solvers in skirting the theoretical intractability of SAT, one wonders,
can these solvers be used to solve set constraints?
We show that this is possible with reasonable performance.
Our full contributions are as follows:
 
\begin{easylist}[itemize]
    & We devise a pattern match analysis for a strict functional language, expressed in terms of unrestricted set constraints (\cref{sec:matchAnalysis}).
    & We provide a method for translating unrestricted set constraint problems into SAT modulo \UF,
    a logical theory with booleans, uninterpreted functions, and first order quantification (\cref{sec:smtTrans}).
    Additionally, we show that projections, a construct traditionally difficult to formulate with set constraints,
    are easily formulated using disjunctions in SMT (\cref{subsec:primer}).  
    & We implement our translation and analysis, showing that they are usable for verification
    despite $\NEXPTIME$-completeness (\cref{sec:impl}).

\end{easylist}

\subsection*{Motivation: Pattern Match Analysis} 
\label{subsec:pattern-match-intro} 

Our primary interest in set constraints is using them to devise a functional \textit{pattern match analysis}.
Many functional programming languages feature \textit{algebraic datatypes}, where values of a datatype $D$ are formed
by applying
a \textit{constructor} function to some arguments. Values of an algebraic type can be decomposed using pattern matching,
where the programmer specifies a number of branches with free variables, and the program takes the first branch that
matches the given value, binding the corresponding values to the free variables. If none of the patterns match the value,
a runtime error is raised.  

Many modern languages, such as Elm~\cite{elmGuide} and Rust~\cite{klabnik2018rust} require that pattern matches be \textit{exhaustive}, so that each pattern match has a branch
for every possible value of the given type. This ensures that runtime errors are never raised due to unmatched patterns,
and avoids the null-pointer exceptions that plague many procedural languages. However, the type systems of these languages cannot express all invariants.
Consider the following pseudo-Haskell, with an algebraic type for shapes, and a function that calculates their area.

\begin{minipage}{0.3\textwidth}
  \begin{haskellCode}
      data Shape = 
        Square Double 
        | Circle Double 
        | NGon [Double]
  \end{haskellCode}
  \end{minipage}
\begin{minipage}{0.65\textwidth}
\begin{haskellCode}
    area :: Shape -> Double
    area shape = case shape of
      NGon sides -> ...
      _ -> simpleArea shape 
      where simpleArea sshape = case sshape of
          Square len -> len * len
          Circle r -> pi * r * r 
          _ -> error "This cannot happen"
\end{haskellCode}
\end{minipage}

The above code is perfectly safe, since \hinline{simpleArea} can only be called from \hinline{area},
and will never be given an \hinline{NGon}. 
However, it is not robust to changes. If we add the constructor \hinline{ Triangle Double Double Double}
to our \hinline{Shape} definition, then both matches are still exhaustive, since the \hinline{_} pattern covers every possible case.
However, we now may face a runtime error if \hinline{area} is given a \hinline{Triangle}.
In general, requiring exhaustiveness forces the programmer to either manually raise an error
or return a dummy value in an unreachable branch.

We propose an alternate approach: remove the catch-all case of \hinline{simpleArea}, and use a static analysis to determine that only
values matching \hinline{Circle} or \hinline{Square} will be passed in. Such analysis would mark the above code safe, but would
signal unsafety if \hinline{Triangle} were added to the definition of \hinline{Shape}. 

The analysis for this particular case is intuitive, but can be complex in general:

\begin{easylist}
    & Because functions may be recursive, we need to be able to handle recursive equations (or inequations)
    of possible pattern sets. For example, a program dealing with lists may generate a constraint of the form
    $X \subseteq Nil \cup Cons(\top, Cons(\top, X))$. 
    & We wish to encode \textit{first-match semantics}: if a program takes a certain branch in the pattern match,
    then the matched value cannot possibly match any of the previous cases.
    & We wish to avoid false negatives by tracking what conditions must be true for a branch to be taken,
    and to only enforce constraints from that branch when it is reachable.
    If we use logical implication,
    we can express constraints of the form ``if $x$ matches pattern $P_1$, then $y$ must match pattern $P_2$''.
\end{easylist}

\cref{sec:matchAnalysis} gives such an analysis, while
\cref{sec:smtTrans} describes solving these constraints.  
Both are implemented and evaluated in \cref{sec:impl}.

\section{A Set Constraint-based Pattern Match Analysis}
\label{sec:matchAnalysis}

Here, we describe an annotated type system for \textit{pattern match analysis}.
It tracks the possible values that expressions may take. %
Instead of requiring that each match be exhaustive, we restrict functions to reject inputs that
may not be covered by a pattern match in the function's body.
 Types are refined by constraints, which are solved using an external solver (\cref{sec:smtTrans}). 

\subsection{\LANG\  Syntax  }

\begin{figure}[t]
  ${x \in \textsc{ProgVariabe},}\ {X \in \textsc{TypeVariable},}\ {D \in
    \textsc{DataType},}\ {K\in\textsc{DataConstructor}}$
  
\begin{minipage}{0.55\textwidth}
\vspace{-12pt} 
\begin{align*}
     &\textbf{Terms} \\
 & t  ::=  x \mid \lambda x \ldotp t  \mid \match t \with \{ \seq{P => t;} \} \\
  &  \qquad \mid t_1\ t_2 \mid K_D(\seq{t}) \mid \llet x = t_1 \iin t_2\\
  &    \textbf{Patterns} \\
  & P  ::=  x \mid  K_D(\seq{P})  \\
  &      \textbf{Underlying Types} \\
  & \tau  ::=  X \mid D \mid \tau_1 -> \tau_2 
\end{align*}
\end{minipage}
\begin{minipage}{0.35\textwidth}
  \begin{align*}
     &    \textbf{Datatype environments} \\
     & \Delta  ::=  \cdot \mid D = \seq{K(\seq{T})}, \Delta\\
     &    \textbf{Underlying Type Schemes} \\
     & \sigma  ::=  \forall\seq{X}\ldotp \tau\\
     &       \textbf{Type Environments} \\
     &    \Gamma  ::=  \cdot \mid X, \Gamma \mid x : T, \Gamma
  \end{align*}
  \end{minipage}
\caption{\LANG: syntax} 
\label{fig:lang-syntax}
\end{figure}

We present \LANG, a small, typed functional language, whose syntax we give in
\cref{fig:lang-syntax}.  Throughout, for a given metavariable $\mathcal{M}$  we write ${\seq{\mathcal{M}}}^i$ for a sequence of objects matching
$\mathcal{M}$. We omit the positional index $i$ when it is unneeded.

In addition to functions and applications, we have a form $K_D(\seq{t})$
which applies the data constructor $K$  to the argument sequence $\seq{t}$
to make a term of type $D$.
Conversely, the form $\match t' \with \{ \seq{P => t;} \}$ chooses the first branch
$P_i => t_i;$ for which $t'$ matches pattern $P_i$, and then evaluates $t_i$ after binding the matching parts
of $t'$ to the variables of $P_i$.
We use Haskell-style shadowing for non-linear patterns: e.g. $(x,x)$ matches any pair, and binds the
second element to $x$. We omit advanced matching features, such as guarded
matches,
since these can be desugared into nested simple matches.
We use type environments $\Gamma$ store free type variables and types for
program variables.
We assume a fixed datatype environment $\Delta$ that stores the names of each datatype $D$, along with the name and argument-types of
each constructor of $D$. 

\subsection{The Underlying Type System}

The underlying type system is in the style of \citet{Damas:1982:PTF:582153.582176},
where monomorphic types are separated from polymorphic type schemes.
The declarative typing rules for the underlying system 
are standard (\cref{fig:decl-types}).
We do not check the exhaustiveness of matches, as
this overly-conservative check is precisely what we aim to replace.
The analysis we present below operates on these
underlying typing derivations, so each expression has a known underlying type.

  \begin{figure}[t]
  \begin{boxedarray}{@{}l@{}}
    \boxed {\Gamma  |- t : \tau}  \text{ (Expression typing)} \\
    \begin{inferbox}
      \inference[\textsc{Ctor}]
      {K(\seq{\tau}) \in \Delta(D)\\
        \seq{ \Gamma  |- t : \tau }
      }{\Gamma  |- K_D(\seq{t}) : D} 

        \inference[\textsc{Lam}]
        {x:\tau_1, \Gamma  |- t : \tau_2
        }{\Gamma  |- \lambda x \ldotp t : \tau_1 -> \tau_2} 

        \inference[\textsc{App}]
        {\Gamma  |- t_1 : \tau_1 -> \tau_2\\
        \Gamma  |- t_2 : \tau_1
        }{\Gamma  |- t_1\ t_2 : \tau_2} 

        \inference[\textsc{Var}]{\Gamma(x) = \forall\seq{X}\ldotp\tau \\
        }{\Gamma  |- x : \seq{[\tau' / X]}\tau}  

        \inference[\textsc{Mat}]
        {
        \Gamma  |- t : \tau\and
        \seq{ \Gamma  |- P : \tau | \Gamma' }\and
        \seq{\Gamma'  |- t' : \tau'}
        }{\Gamma  |- \match t \with \{ \seq{P => t'}; \} : \tau' } 

        \inference[\textsc{Let}]
        { x : \tau_1, \seq{X}, \Gamma  |- t_1 : \tau_1  \and
          x : \forall \seq{X} \ldotp \tau_1, \Gamma  |- t_2 : \tau_2
        }{\Gamma  |- \llet x = t_1 \iin t_2 : \tau_2 } 
        
    \end{inferbox}\\\\  

    \boxed {\Gamma  |- P : \tau | \Gamma'} \text{ (Pattern typing and binding generation)}\\
    \begin{inferbox}
      \inference[\textsc{Var}]{
      }{\Gamma  |- x : \tau | (x : \tau),\Gamma }

        \inference[\textsc{Ctor}]
        {K(\seq{\tau}) \in \Delta(D)\and 
          \seq{ \Gamma  |- P : \tau | \Gamma' }
        }{\Gamma  |- K_D(\seq{P}) : D | \bigcup{\seq{\Gamma'}}}

    \end{inferbox}\\\\

  \end{boxedarray}\\
  \caption{Underlying Typing for Expressions and Patterns}
  \label{fig:decl-types}
\end{figure}

\subsection{Annotated Types}

\begin{figure}[t]
  \vspace{-20pt}  
  \begin{minipage}{0.45\textwidth}
\begin{align*}
  &V &  \in & && \textsc{SetVariable}\\ 
  &&  &&& \textbf{Set constraints} \\
  &C &  ::= & &&E_1 \subseteq E_2 \mid C_1 \land C_2 \mid C_1 \lor C_2 \mid \lnot C\\ 
  &&  &&& \textbf{Annotated Types} \\
  &T & ::=  &&& X^E \mid D^E \mid  (T_1 -> T_2)^E
  \end{align*}
\end{minipage}
\begin{minipage}{0.45\textwidth}
  \begin{align*}
    &&  &&& \textbf{Set expressions} \\
    &E & ::= &&& V \mid E_1 \cup E_2 \mid E_1 \cap E_2 \mid \lnot E \\
    &&&&& \mid K_D(\seq{E}) \mid  K_D^{-i}(E) \mid \top \mid \bot\\
    &&  &&& \textbf{Annotated Schemes} \\
    &S & ::= &&& \forall\seq{X}, \seq{V}\ldotp C => T
    \end{align*}
  \end{minipage}
\caption{\LANG: annotations} 
\label{fig:lang-annot}
\end{figure}

For our analysis, we annotate types with \textit{set expressions}
(\cref{fig:lang-annot}).
We define their semantics formally in \cref{sec:smtTrans}, but intuitively,
they represent
possible shapes that the value of an expression might have in some context.
We have variables, along with intersection, union and negation,
and $\top$ and $\bot$ representing the sets of all and no values respectively.
The form $K_D(E_1 \ldots E_a)$ denotes applying the arity-$a$ constructor $K$ of datatype $D$
to each combination of values from the sets denoted by $\seq{E}$ .
Conversely, $  K_D^{-i}(E) $ denotes the $i$th \textit{projection} of $K$: it takes the $i$th argument of each value constructed
using $K$ from the set denoted by $E$.

\textit{Set constraints} then
specify the inclusion relationships between those sets.
These are boolean combinations of atomic constraints of the form $E_1 \subseteq
E_2$.
Our analysis uses these in \textit{annotated type schemes}, to constrain which annotations
a polymorphic type accepts as instantiations. The idea is similar to Haskell's
typeclass constraints, and we adopt a similar notation.  
Since each syntactic variant has a top-level annotation $E$, we use $T^E$ to denote
an annotated type $T$ along with its top-level annotation $E$.
Annotated types $T^E$ replace underlying types $\tau$ in our rules, and our analysis emits constraints on $E$
that dictate its value.
We note that boolean operations such as $\implies$ and $\iff$, 
can be decomposed into $\land$, $\lor$, and $\lnot$. Similarly, we use
$E_1 = E_2$ as a shorthand for $E_1 \subseteq E_2 \land E_2 \subseteq E_1$,
and  $\true$ and $\false$ as shorthands for $\bot\subseteq\top$ and
$\top\subseteq\bot$ respectively.

\subsection{The Analysis}

  \begin{figure}[t!]
  \begin{boxedarray}{@{}l@{}}
    
    \boxed{\Gamma  | C_p |- t : T_E  \withConstr C }  \text{ (Pattern Match Analysis) }\\
    \begin{inferbox}
      \inference[\textsc{AVar}]{\Gamma(x) = \forall\seq{X},\seq{V}\ldotp C => T^E \\
      \seq{V'} \textbf{ fresh }
    }{\begin{array}{@{}c@{}}
      \Gamma  | C_p |- x : \seq{[V' / V]}\seq{[\tau' / X]}(T^E )\\ \withConstr (C_p
        \implies \seq{[V' / V]}C)
      \end{array}}

    \inference[\textsc{AApp}]
    {\Gamma  | C_p |- t_1 : (T_1^{E_1} -> T_2^{E_2})^{E_3} \withConstr C_1\\ 
      \Gamma  | C_p |- t_2 : T_1^{E'_1} \withConstr C_2
    }{\begin{array}{@{}c@{}}
        \Gamma  | C_p |- t_1\ t_2 : T_2^{E_2} \\ \withConstr\! C_1 \!\land\! C_2 \!\land\! (C_p
        \!\!\implies\!\! T_1^{E_1} \equiv T_1^{E'_1})
      \end{array}} 

    \inference[\textsc{ACtor}]
    {K(\seq{T}) \in \Delta(D)\\ 
      \seq{ \Gamma  | C_p |- t : T^E \withConstr C}\\
    }{\Gamma  | C_p |- K_D(\seq{t}) : D^{K(\seq{E})} \withConstr    \bigwedge \seq{C}}   

        \inference[\textsc{ALam}]
        { V \textbf{ fresh }\\
          x:T_1^{V}, \Gamma  | C_p |- t : T_2^{E} \withConstr C
        }{\Gamma  | C_p |- \lambda x \ldotp t \!:\! (T_1^{V} \!\!->\!\! T_2^{E})^\top
          \withConstr\! C }

        \inference[\textsc{AMat}] 
        {
          V \textbf{ fresh }\and \Gamma  | C_p |- t : T^E \withConstr C_{dsc}  \and
          \seq{\Gamma  | C_p |- P_i : T^{E \cap \mathcal{\overline{P}}_i(\seq{P}) } | \Gamma_i }^i\\
          \seq{C_i  := (E  \cap \Pb{P_i} \cap \mathcal{\overline{P}}_i(\seq{P})
          \not\subseteq \bot) }^i\and
          \seq{\Gamma_i  | C_{i} \land C_p |- t'_i :  T'^{E'_i}}^i \\
        C_{res} := \bigwedge \seq{ C_i \implies  E'_i \subseteq V}^i \and
         C_{saf} := (C_p \implies (E \subseteq \bigcup \seq{ \Pb{P_i}}^i)) 
        }{\begin{array}{@{}c@{}}
          \Gamma  | C_p |- \match t \with \{ \seq{P_i => t'_i} \} : T'^V 
           \withConstr C_{dsc} \land C_{res} \land C_{saf} 
        \end{array} } 

        \inference[\textsc{ALet}]
        { {T'}_1^{V'} := \textbf{freshen}(T_1) \and
          x : T'_1, \seq{X}, \Gamma  | C_p |- t_1 : T_1^E \withConstr C_1  \\ 
          \seq{V} = (\FV(E) \cup \FV(C_1)) \setminus (\FV(\Gamma) \cup \FV(C_p)) \and
          {T'}_1^{V'} \equiv T_1^E \land C_1 \text{ satisfiable } \\
          x : (\forall \seq{X}, \seq{V} \ldotp ({T'}_1^{V'} \equiv T_1^E \land C_1) => T_1^E), \Gamma  | C_p |- t_2 : T_2^{E_2} \withConstr C_2
        }{\Gamma  | C_p |- \llet x = t_1 \iin t_2 :  T_2^{E_2} \withConstr C_2 } 
        
    \end{inferbox}\\\\  

    \boxed {\Gamma  |- P : T^E | \Gamma' } \text{ (Analysis pattern environments. $P, T^E$ are input, $\Gamma'$ is output)}\\
    \begin{inferbox}
      \inference[\textsc{Var}]{
      }{\Gamma  |- x : T^E | (x : T^E),\Gamma } 

        \inference[\textsc{Ctor}]
        {K(T_1,\ldots,T_n) \in \Delta(D) \\
          {\seq{\Gamma  |- P : T^{K^{{-i}}(E)} : \Gamma_1' }}^i
        }{\Gamma  |- K_D(\seq{P}) : D^E | \bigcup{\seq{\Gamma'}}^i} 
      \end{inferbox}
\\\\  
  \end{boxedarray}
  \caption{Pattern Match Analysis}
  \label{fig:the-analysis}
\end{figure}

\begin{figure}
  \fbox{\begin{minipage}{\textwidth}
\begin{align*}
  &\boxed{T_1 \equiv T_2 := C} \qquad \text{(Type equating)}\\
  &  (T_1 -> T'_1)^{E_1} \equiv (T_2 -> T'_2)^{E_2} := (T_1 \equiv T_2) \land (T'_1 \equiv
  T'_2) \land E_1 = E_2 \\
  &X^{E_1} \equiv X^{E_2} := E_1 = E_2 \qquad D^{E_1} \equiv D^{E_2} := E_1 = E_2 \qquad
  T_1 \equiv T_2 := \false \textit{ otherwise} \\
  &\boxed{\textbf{freshen}(T) := T} \qquad \text{(Annotation freshening where $V$ fresh)}\\
  &  \textbf{freshen}(X^E) := X^V \qquad \textbf{freshen}(D^E) := D^V \\
  &\textbf{freshen}((T_1 -> T_2)^E) := (\textbf{freshen}(T_1) ->
  \textbf{freshen}(T_2))^V\\
  & \boxed{\Pb{x} := E} \text{ (Set expression matched by pattern )}\\
  &\Pb{x}  := \top \qquad \Pb{K(\seq{P})} := K(\seq{\Pb{P}})\\
  & \boxed{\mathcal{\overline{P}}_i(\seq{P}) := C} \qquad \text{(Not-yet covered pattern at branch $i$)}\\
  &\mathcal{\overline{P}}_0(P_0 \ldots P_n) = \top \qquad \mathcal{\overline{P}}_i(P_0 \ldots P_n) = \lnot \Pb{P_0} \cap \ldots \lnot
    \Pb{P_{i-1}} \text{ when $0 < i \leq n$}
\end{align*}
\end{minipage}}
\caption{Auxiliary Metafunctions}
\label{fig:aux-metafun}
\end{figure}

We present our pattern match analysis in \cref{fig:the-analysis}. 
The analysis is phrased as an annotated type system in the style of \citet{Nielson:1999:TES:646005.673740}.
The judgment $\Gamma | C_p |- t : T^E \withConstr C$
says that, under context $\Gamma$,
if $C_p$ holds, then
$t$ has the underlying type of $T$ and can take only forms from $E$, where
the constraint $C$ holds.
$C_p$ is an input to the judgment called the \textit{path constraint},
which must hold for this part of the program to have been reached.
The set expression $E$ and constraint $C$ are outputs of the judgment,
synthesized by traversing the expression.
We need an external solver for set constraints
to find a value for each variable $V$ that satisfies $C$.
This is precisely what we define in \cref{sec:smtTrans}.
We write the conversion between patterns and set-expressions as
$\Pb{P}$.

The analysis supports higher-order functions, and it is \textit{polyvariant}:  refined types use polymorphism,
so that precise analysis can be performed at each instantiation site.
A variant of Damas-Milner style inference with let-generalization is used to generate these
refined types.
Moreover, the analysis is push-button: no additional input need be provided
by the programmer.
It is sound but conservative: it accounts for all possible values
an expression may take, but may declare  some matches unsafe
when they will not actually crash.
The lack of polymorphic recursion is a source of imprecision,
but a necessary one for preserving termination without
requiring annotations from the programmer.

We generate two sorts of constraints. First, we constrain what values expressions could possibly take.
For example, if we apply a constructor $K_D(\seq{t})$,
and we know the possible forms $\seq{E}$ for $\seq{t}$, then in any context,
this expressions can only ever evaluate to values in the set $K(\seq{E})$.
Second, we generate safety constraints, which must hold to ensure that the program encounters
no runtime errors. Specifically, we generate a constraint that when we match on a term $t$,
all of its possible values are covered by the left-hand side of one of the branches.

{\textbf{Variables:} }
Our analysis rule $\textsc{AVar}$ for variables looks up a scheme from $\Gamma$.
However, typing schemes now quantify over type and set variables, and carry a constraint
along with the type. We then take instantiation of type variables as given,
since we know the underlying type of each expression.
Each set variable is instantiated with a fresh variable.
We then give $x$ the type from the scheme, with the constraint that the instantiated
version of the scheme's constraint must hold if this piece of code is reachable
(i.e. if the path condition is satisfiable). 

{\textbf{Functions and Applications:}}
The analysis rule $\textsc{ALam}$ for functions is straightforward. We generate a fresh set variable
with which to annotate the argument type in the environment, and check the body
in this extended environment. Since functions are not algebraic datatypes
and cannot be matched upon, we emit $\top$ as a trivial set of possible
forms for the function itself.

We know nothing about the forms that the parameter-type annotation $V$ may take, since it depends entirely
on what concrete argument is given when the function is applied.
However, when checking the body, we may encounter a
pattern match that constraints what values $V$ may take without risking runtime failure.
So our analysis may emit safety constraints involving $V$, but it will not constrain it otherwise.
Generally, $(T_1^{E_1} -> T_2^{E_2})$ means that the function can safely accept
any expression matching $E_1$, and may return values matching $E_2$.

Applications are analyzed using $\textsc{AApp}$. %
Annotations and constraints for the function and argument are both generated,
and we emit a constraint equating the argument's annotated type with its domain, under the assumption that
the path condition holds and this
function call is actually reachable.
The metafunction $T_1^{E_1} \equiv T_1^{E'_1}$ (defined in \cref{fig:aux-metafun}) traverses the structure of the argument and function domain type,
constraining that parallel annotations are equal. This traversal is possible because the underlying
type system guarantees that the function domain and argument have identical underlying types.  

{\textbf{Constructors:}}
As we mentioned above, applying a constructor to arguments can only produce a value that is that constructor wrapped around
its argument's values. The rule $\textsc{ACtor}$ for a constructor $K$ infers annotations and constraints for each argument,
then emits those constraints and applies $K$ to those annotations.

{\textbf{Pattern matching:}}
It is not surprising that in a pattern match analysis, the interesting details
are found in the case for pattern matching.
The rule $\textsc{AMat}$ begins by inferring the constraint $C_{dsc}$ and
annotation $E$ for
the discriminee $t$. 

For each branch, we perform two tasks.
First, for each branch's pattern $P_i$, we use an auxiliary judgment to generate the environment $\Gamma_i$ binding the pattern variables
to the correct types and annotations, using projection to access the relevant parts.
For $P_1$, the annotation of the whole pattern is $E$ i.e. the annotation for $t$. However, the first-match
semantics mean that if we reach $P_i$, then the discriminee does not match any of
$P_1 \ldots P_{i-1}$. So for each $P_i$, we extend the environment with annotations obtained by intersecting $E$
with the negation of all previous patterns, denoted 
$\mathcal{\overline{P}}_i(\seq{P})$ (\cref{fig:aux-metafun}).

Having obtained the extended environment for each branch,
we perform our second task: we check each right-hand-side in the new
environment, obtaining an annotation $E'_i$.
When checking the results, we augment the
path constraint with $C_i$, asserting that some possible input
matches this branch's pattern, obtained via $\Pb{}$ (\cref{fig:aux-metafun}), but
none of the previous.
This ensures that safety constraints
for the branch are only enforced when the branch can actually be taken.

To determine the annotation for the entire expression, we could naively take the union of the annotations
for each branch. However, we can be more precise than this.
We generate a fresh variable $V$ for the return annotation, and constrain
that it contains the result $E'_i$ of each branch, provided
that it $C_i$ holds, and it is possible we actually took that branch. 
This uses implication, justifying the need for a solver
that supports negation and disjunction.

Finally, we emit a safety constraint $C_{saf}$, saying that if it is possible to reach this part of the program (that is, if $C_p$ holds),
then the inputs to the match must be contained within
the values actually matched.

{\textbf{Let-expressions:}}
Our \textsc{ALet} rule deals with the generalization of types into type schemes.
This rule essentially performs Damas-Milner style inference, but for the annotations, rather
than the types. 
When defining $x = t_1$, we check $t_1$
in a context extended with its type variables, and a monomorphic version of its own type.
The metafunction $\mathbf{freshen}$ takes the underlying type for $t_1$ and adds fresh annotation variables
across the entire type. This allows for monomorphic recursion.
The metafunction $\equiv$ constrains the freshly generated variables
on $T'$ to be equal to the corresponding annotations
on $T_1$ obtained when checking $t_1$.
Again this traversal is possible because the underlying types must be identical.
Once we have a constraint for the definition, we check that its constraint is in fact satisfiable,
ensuring that none of the safety constraints are violated. In our implementation, this is where
the call to the external solver is made.

To generate a type scheme for our definition, we generalize over all variables free in the inferred annotation
or constraint but not free in $\Gamma$ or $C_p$.
Finally, we check the body of the let-expression in a context extended with the new variable and type scheme.
Because  let-expressions are where constraints are actually checked, we assume that all top-level definitions
of a program are wrapped in let-declarations, and are typed with environment $\cdot$ and path constraint $\true$.

\textbf{Example -  Safety Constraints:}
To illustrate our analysis, we return to the \hinline{Ngon} code from \cref{subsec:pattern-match-intro}.
We assume that all \hinline{Double} terms are given annotation $\top$.
Then, the \hinline{simpleArea} function would be given the annotated type scheme
$\forall V_1 , V_2 \ldotp C_1 \land C_2 \land C_3 => \mathtt{Ngon}^{V_1} \to
\mathtt{Double}^{V_2}$, where
\begin{gather*}
   C_1 :=V_1 \subseteq \mathtt{Square}(\top) \cup \mathtt{Circle}(\top) \quad
   C_2 := (V_1 \cap \mathtt{Square}(\top) \not\subseteq \bot) \implies  \top \subseteq V_2 \\
   C_3 := ((V_1 \cap \mathtt{Circle}(\top) \cap \lnot \mathtt{Square}(\top)) \not\subseteq \bot) \implies  \top \subseteq V_2 
 \end{gather*}
 $C_1$ is the $C_{saf}$ generated by the \textsc{AMat} rule,
 saying that the function can safely accept input from ${\mathtt{Square}(\top) \cup \mathtt{Circle}(\top)}$.
 $C_2$ and $C_3$ are conjuncts of $C_{res}$, describing how,
 if the input overlaps with $\mathtt{Square}$ then the output can be anything,
 and that if the input overlaps with $\mathtt{Circle}$ but not
 $\mathtt{Square}$, then the output can be anything.
 $C_2$ and $C_3$ are trivially satisfiable: $\mathtt{Circle}(\top) \cap
 \lnot\mathtt{Square}(\top)$ is  $\mathtt{Circle}(\top)$,
 so they are essentially saying that $V_2$ must be $\top$.

 When we call \hinline{simpleArea} from \hinline{area}, we are in the branch after the \hinline{Ngon} case has been checked.
The scheme for \hinline{simpleArea} is instantiated with the path constraint
$V_4 \subseteq \top \cap \lnot (\mathtt{Ngon}(\top)) $, where $V_4$ is the annotation for \hinline{shape},
because it is called after we have a failed match with \hinline{Ngon sides}.

Suppose we instantiate $V_1, V_2$ with fresh $V'_1, V'_2$.
The call to \hinline{simpleArea} creates a constraint that $V_4 = V'_1$.
Taking this equality into account, the safety constraint
is instantiated to $ V_4 \subseteq \top \cap \lnot (\mathtt{Ngon}(\top)) \implies V_4  \subseteq (\mathtt{Square}(\top) \cup \mathtt{Circle}(\top))$.
This is satisfiable for any value of \hinline{shape}, so at every call to \hinline{area}
the analysis sees that the safety constraint is satisfied. If we add a
$\mathtt{Triangle}$ constructor, then the constraint is unsatisfiable
any time $V_4$ is instantiated to a set with $\mathtt{Triangle}$. 

\textbf{Example - Precision on results of matching:}
To illustrate the precision of our analysis for the \textit{results} of pattern matching,
we turn to a specialized version of the classic \hinline{map} function:

\begin{haskellCode} 
intMap : (Int -> Int) -> List Int -> List Int ->
intMap f l = case l of
  Nil -> Nil
  Cons h t -> Cons (f h) (intMap f t)    
\end{haskellCode}

Suppose we have concrete arguments $\mathtt{f} : (\mathtt{Int}^{V_{11}} -> \mathtt{Int}^{V_{12}})^{V_1}$ and $\mathtt{l} : (\mathtt{List Int})^{V_2}$.
The safety constraint for the match is that $V_2 \subseteq \mathtt{Nil} \cup \mathtt{Cons}(\top, \top)$,
which is always satisfiable since the match is exhaustive. The result of the case expression is given a fresh variable annotation
$V_3 $.  From the first branch,
we have the constraint that $V_2 \cap \mathtt{Nil} \not\subseteq \bot \implies \mathtt{Nil} \subseteq V_3 $.

The analysis is more interesting for the second branch.
The bound pattern variables $h$ and $t$ are given annotations $\mathtt{Cons}^{-1}(V_2)$ and  $\mathtt{Cons}^{-2}(V_2)$
respectively, since they are the first and second arguments to $\mathtt{Cons}$.
 Because our recursion is monomorphic, 
 the recursive call \hinline{intMap f t} generates the trivial constraint $ {(V_2 \cap \lnot\mathtt{Nil} \cap \mathtt{Cons}(\top, \top)) \not\subseteq \bot \implies V_1 \subseteq V_1}$, 
 and the more interesting constraint ${(V_2 \cap \lnot\mathtt{Nil} \cap \mathtt{Cons}(\top, \top)) \not\subseteq \bot \implies \mathtt{Cons}^{-2}(V_2) \subseteq V_2}$.
This second constraint may seem odd, but it essentially means that without polymorphic recursion,
our program's pattern matches must account for any length of list.
This is where having set constraints is extremely useful: if we were to use some sort of symbolic execution
to try to determine a single logical value that $l$ could take, then treating the recursive call monomorphically would create an impossible equation.
But the set $\mathtt{Cons(a,\mathtt{Nil})}, \mathtt{Cons(a,\mathtt{Cons}(b,\mathtt{Nil})), \ldots}$
satisfies our set constraints, albeit in an imprecise way.

When checking the body, suppose that $V_5$ is the fresh variable ascribed
to the return type of \hinline{intMap}.
For the result of the second branch, we have the constraints
$V_2 \cap \lnot\mathtt{Nil} \cap \mathtt{Cons}(\top,\top) \not\subseteq \bot \implies \mathtt{Cons}(V_{12}, V_5) \subseteq V_3$.
This essentially says that if the input to the function can be \hinline{Cons}, then so can the output,
but if the input is always \hinline{Nil}, then this branch contributes nothing to the overall result.
Finally, we have a constraint $V_5 = V_3$, generated by the metafunction $\equiv$.

Our result annotation $V_3$ is constrained by
$(V_2 \cap \mathtt{Nil} \not\subseteq \bot \implies \mathtt{Nil} \subseteq V_3) $\\
 $ \ \land\   (V_2 \cap \lnot\mathtt{Nil} \cap \mathtt{Cons}(\top,\top) \not\subseteq \bot \implies \mathtt{Cons}(V_{12}, V_3) \subseteq V_3)$,
capturing how \hinline{intMap} returns nil empty result for
nil input, and non-nil results for non-nil input.

\section{Translating Set Constraints to SMT} 
\label{sec:smtTrans}

While the above analysis provides a fine-grained way to determine which pattern matches
may not be safe, it depends on the existence of an external solver to check
the satisfiability of the resulting set constraints. 
We provide a simple, performant solver by translating
set constraints into an SMT formula.
\subsection{A Primer in Set Constraints} 
\label{subsec:primer}

We begin by making precise the definition of the set constraint problem.
Consider a set of (possibly 0-ary) functions $\F = \{f^{a_1}_1, \ldots, f^{a_n}_n\}$, where each $a \geq 0$
is the arity of the function $f^a_i$.
The \textit{Herbrand Universe} $\H_\F$ is defined inductively: each $f^0_i \in \F$ is in $\H_\F$, and if $a > 0$ and $h_1, \ldots, h_a$ are in $\H_\F$,
then $f^a_i(h_1, \ldots, h_a)$ is in $\H_\F$.
(We write $\H_\F$ as $\H$ when the set $\F$ is clear.)
Each $f^a_i$ is injective, but is otherwise uninterpreted, behaving like a constructor
in a strict functional language.
We assume all terms are finite, although similar analyses
can account for laziness and infinite data~\citep{Koot:2015:TEA:2678015.2682542}.

\begin{figure}[t]
  \begin{minipage}{0.45\textwidth}
    \begin{align*}
    \Hb{\setbot }{\sigma} & = & \emptyset \\
    \Hb{\top }{\sigma} & = & \H \\
    \Hb{V }{\sigma} & = & \sigma(V) \\ 
      \Hb{\lnot{E_1} }{\sigma} & = & \H \setminus \Hb{ E_1 }{\sigma}\\
      \end{align*}
  \end{minipage}
  \hfill\vline\hfill
  \begin{minipage}{0.45\textwidth}
    \begin{align*}
    \Hb{E_1 \cap E_2 }{\sigma} & = & \Hb{ E_1 }{\sigma} \cap \Hb{ E_2 }{\sigma}  \\
    \Hb{E_1 \cup E_2 }{\sigma} & = & \Hb{ E_1 }{\sigma} \cup \Hb{ E_2 }{\sigma}  \\
    \Hb{f_i^a(E_1, \ldots, E_a)}{\sigma} & = & \{ f_i^a(h_1, \ldots, h_a)  \\
      &&  \mathllap { \mid h_1 \in \Hb{E_1}{\sigma}, \ldots,  h_a \in \Hb{E_a}{\sigma} \} }
    \end{align*}
    \end{minipage}
    \caption{Semantics of Set Expressions}
    \label{fig:setExpGrammar}
  \end{figure}

  This allows us to formalize the semantics of set expressions.
  The syntax is the same as in\cref{fig:lang-annot},
  although we use the notation $f^a_i(\seq{E})$ instead of $K_D(\seq{E})$
  to denote that we are using arbitrary function symbols
  from some Herbrand universe $\H$,
  instead of specific constructors for a datatype.
Given a substitution ${\sigma : \V -> \P(\H)}$, we can assign a meaning $\Hb{E}{\sigma} \subseteq \H$ for an expression $E$
by mapping variables to their substitutions, and applying the corresponding set
operations.
The full semantics are given in \cref{fig:setExpGrammar}.
Note that the expressions on the left are to be interpreted as \textit{syntax},
whereas those on the right are mathematical sets.

A \textit{set constraint atom} $\mathcal{A}$ is a constraint of the form $E_1 \subseteq E_2$. 
These are also referred to as \textit{positive set constraints} in previous work. 
A \textit{set constraint literal} $\L$ is either an atom or its negation $\lnot (E_1 \subseteq E_2)$, which we write as $E_1 \not\subseteq E_2$.

Constraints which contain negative literals are called \textit{negative set
  constraints}.
An unrestricted set constraint, denoted by metavariable $\C$, is a boolean combination (i.e. using
$\land$, $ \lor$ and $\lnot$) of set constraint atoms, as we defined in \cref{fig:lang-annot}.
For example, $(X \subseteq Y \implies Y \subseteq X)\land (Y \not\subseteq Z) $ is an unrestricted set constraint. 

Given a set constraint $C$, the satisfiability problem is to determine whether there exists
a substitution ${\sigma : \V -> \P(\H)}$ such that, if each atom $E_1 \subseteq E_2$ in $C$ 
is replaced by the truth value of $\Hb{E_1}{\sigma} \subseteq \Hb{E_2}{\sigma} $, then 
the resulting boolean expression is true.
Since solving for arbitrary boolean combinations of set constraints is difficult, we focus on a more restricted version
of the problem. The \textit{conjunctive} set constraint problem for a sequence of literals $\seq{L}$ is to find a variable assignment that causes
 $\bigwedge\seq{L}$ to be true.
 We explain how to extend our approach to arbitrary boolean combinations in \cref{subsec:bool-comb}.

One can see that the Herbrand universe $\H$ closely matches the set of terms that can be formed
from a collection of algebraic datatypes, and that allowing negative constraints and arbitrary boolean expressions satisfies
the desiderata for our pattern match analysis. 

\subsection{Projection} 
Many analyses (including ours) on a notion of \textit{projection}. For a set expression $E$,
we denote the $j$th projection of $E$ for  function $f_i^a$ by $f_i^{-j}(E)$.
For a substitution $\sigma$,
we have $\Hb{f_i^{-j}(E)}{\sigma} = \{ h_j \mid f^a_i(h_1, \ldots, h_j,  \ldots h_a) \in E  \}$.

While we don't explicitly include projections in our grammar for set expressions, we 
can easily express them using boolean formulae. Given some constraint $C[f_i^{-j}(E)]$,
we can replace this with:\\
${C[X_j] \land (E\cap f_i^a(\top,\ldots,\top)) = f_i^a(X_1, \ldots, X_j, \ldots X_a) \land (E = \setbot \iff X_j = \setbot )}$\\
 where each $X_k$ is a fresh variable.
The first condition specifies that our variable holds the $j$th component of
every $f(\seq{h})$ in $E$.
The second condition is necessary because $f_i^a(X_1, \ldots, X_j, \ldots X_a) = \setbot$ if \textit{any} $X_k$
is empty, so any value of $X_j$ vacuously satisfies $E' = f_i^a(X_1, \ldots, X_j, \ldots X_a)$ if $E'$ and some $X_i$ are empty.

\subsection{Set Constraints and Monadic Logic}

The first step in our translation is converting a conjunction of set constraint literals into a formula in first-order monadic logic,
for which satisfiability is decidable. We then translate this into a search for a solution to 
an SMT problem over $\UF$, the theory of booleans, uninterpreted functions
and first-order quantification.
We gradually build up our translation, first translating set constraints into monadic logic, then translating
monadic logic into SMT, then adding optimizations for efficiency. The complete translation is given in
\cref{sec:complete-trans}. 

Monadic first order logic, sometimes referred to as the \textit{monadic class},
consists of formulae containing only unary predicates, boolean connectives, and constants.
\citet{287598} found a translation from a conjunction $\bigwedge\seq{L}$ of positive set constraint atoms to an equisatisfiable
monadic formula,
which was later extended to negative set constraints with equality~\citep{316078}.
We summarize their procedure here, with a full definition in \cref{fig:monadicTranslation}.
For each sub-expression $E$ of $\bigwedge \seq{L}$, we create a predicate $P_E(x)$,
denoting whether an element $x$ is contained in $E$.
Along with this, the formula $\Eb{E}$ gives the statement that must hold for $P_E$ to respect the semantics of set expressions. 
This is similar to the Tsieten transformations used to efficiently convert arbitrary formulae to a normal form~\cite{Tseitin1983}.
Given $P_E(x)$ for each $E$, we can represent the constraint $E_1 \subseteq E_2$ as ${\forall x \ldotp (P_{E_1}(x) \implies P_{E_2}(x))}$.
Similarly, $E_1 \not\subseteq E_2$ corresponds to ${\exists x \ldotp (P_{E_1}(x) \land \lnot P_{E_2}(x))}$.
\footnote{
The original translation transformed constants and functions into existential variables.
We skip this, since SMT supports uninterpreted functions and constants.}

\begin{figure}[t]
  $M \in \M$ (Monadic formulae)\\ 
  $\boxed{ \Eb{E} = M} $ (Predicates for set expressions)
  \begin{align*}
    &\mathrlap{\Eb{\top}  =  \forall x \ldotp P_\top(x) \qquad|\qquad
      \Eb{\setbot}  =  \forall x \ldotp \lnot P_\setbot(x) \qquad|\qquad
      \Eb{X}  =   \true} &&&&\\
  &\Eb{E_1 \cap E_2} & = &&& \forall x \ldotp P_{E_1 \cap E_2}(x) \iff (P_{E_1}(x) \land P_{E_2}(x))  \\
  &\Eb{E_1 \cup E_2} & = &&& \forall x \ldotp P_{E_1 \cup E_2}(x) \iff (P_{E_1}(x) \lor P_{E_2}(x))  \\
  &\Eb{ \lnot{E_1}} & = &&& \forall x \ldotp P_{\lnot{E_1}}(x) \iff \lnot P_{E_1}(x)  \\
  &\Eb{f^a_i(E_1, \ldots, E_a)} & = &&& ( \forall x_1 \ldots x_a \ldotp P_{f^a_i(E_1, \ldots, E_a)}(f^a_i(x_1, \ldots, x_a)) \iff P_{E_1}(x_1) \land \ldots P_{E_a}(x_a))\\
  &&&&&  (\bigwedge_{g^{a'}_j \neq f^a_i} \forall x_1 \ldots x_{a'} P_{f^a_i(E_1, \ldots, E_a)}(g^{a'}_j (x_1, \ldots, x_a')) \iff \false)   
  \end{align*}
  \begin{minipage}{0.47\textwidth}
    $ \boxed{ \Lb{L} = M}$ (Literal predicates)
  \begin{align*}
  &\Lb{E_1 \subseteq E_2}\  =\  \forall x \ldotp P_{E_1}(x) \implies P_{E_2}(x)  \\
  &\Lb{E_1 \not\subseteq E_2}\  =\ \exists y \ldotp P_{E_1}(y) \land \lnot P_{E_2}(y)  
  \end{align*} 
\end{minipage}
\vrule\ 
\begin{minipage}{0.43\textwidth}
  $\boxed {\Lvb{\bigwedge\seq{L}} = M}$ (Conjunction) 
  \begin{align*}
  &\Lvb{\bigwedge\seq{L}} \ = \  \Eb{E_1} \land \ldots \land \Eb{E_n} \land \bigwedge \seq{\Lb{L} }    \\
  & \text{ where } E_1 \ldots E_n \text{ all subexpressions of }\seq{L}
  \end{align*}
\end{minipage}

    \caption{Translating Set Constraints to Monadic Logic}
    \label{fig:monadicTranslation}
  \end{figure}

The key utility of having a monadic formula is the \textit{finite model property}~\citep{Lowenheim1915, Ackermann}:

\begin{theorem}
  \label{thm:monadic}
    Let $\mathcal{T}$ be a theory in monadic first-order logic with $N$ predicates.
    Then, for any sentence $\mathcal{S}$ in $\mathcal{T}$, there exists a model satisfying
    $\mathcal{S}$ if and only if there exists a model satisfying $\mathcal{S}$ with
    a finite domain of size at most $2^N$.
  \end{theorem}

The intuition behind this is that if there exists a model satisfying $\mathcal{S}$, then
then we can combine objects that have identical truth values for each predicate.
This is enough to naively solve set constraints: we convert them into formulae monadic logic, then search the space of all models
of size up to $2^N$ for one that satisfies the monadic formulae.
However, this is terribly inefficient, and disregards much of the information we
have from the set constraints.

\textbf{Example - Translation:}
Consider $C_3$ from the safety constraint example in \cref{sec:matchAnalysis}.
We see that $\Lb{((V_1 \cap\mathtt{Circle}(\top) \cap \lnot
  \mathtt{Square}(\top)) \not\subseteq \bot) \implies\top \subseteq V_2 }$ is \\
$( \exists y \ldotp P_{V_1 \cap\mathtt{Circle}(\top) \cap \lnot
  \mathtt{Square}(\top)}(y) \land \lnot\false) ==> (\forall x \ldotp\true ==>
P_{V_2}(x))$.
Applying the $\Eb{}$ equivalences for $\cap,\cup$ and $\lnot$ with basic laws of predicate logic
gives us:
$(\exists y \ldotp P_{V_1}(y) \land P_{\mathtt{Circle}(\top)}(y) \land \lnot
P_{\mathtt{Square}(\top)}(y)) ==> \forall x \ldotp P_{V_2}(x) $.\\
Finally, adding the $\Eb{}$ conditions for functions gives us: \\
      $ (\forall x \ldotp P_{\mathtt{Circle}(\top)}(f_ {Circle}(x))  ) \quad \land
      \quad (\forall x \ldotp\lnot  P_{\mathtt{Circle}(\top)}(f_ {Square}(x))  )$\\
      $ \land \quad (\forall x \ldotp P_{\mathtt{Square}(\top)}(f_ {Square}(x)) ) \quad\land\quad (\forall x \ldotp\lnot  P_{\mathtt{Square}(\top)}(f_ {Circle}(x))  )$\\
      $ \land \quad((\exists y \ldotp P_{V_1}(y) \land P_{\mathtt{Circle}(\top)}(y) \land \lnot
      P_{\mathtt{Square}(\top)}(y)) \quad==>\quad \forall x \ldotp P_{V_2}(x)) $.\\
      $C_3$ is satisfiable iff there is a
      model defining predicates $P_{V_1}, P_{V_2}, P_{\mathtt{Circle}(\top)}$ and
      $P_{\mathtt{Square}(\top)}$, and functions $f_{Circle}$,$f_{Square}$
      in which the above formula is true.

\subsection{Monadic Logic in SMT}

To understand how to translate monadic logic into SMT, we first look at what exactly a model
for a monadic theory is. Suppose $\B = \{ \true, \false \}$ is the set of
booleans, which we call \textit{bits}, and say a bit is set if it is $\true$.
For our purposes, a model consists of a set $D$, called the \textit{domain},
along with \textit{interpretations} $I_P : D -> \B$ for each predicate $P$ and $f^a_i : D^a -> D$ for each function, which define the value of $P(x)$ and $f(x_1,\ldots,x_a)$ for each $x,x_1,\ldots,x_a \in D$.
A naive search for a satisfying model could guess $M \leq 2^N$, set $D = \{1 \ldots M \}$, and iterate through all possible truth assignments for each $I_P$,
and all possible mappings for each $f^a_i$,
searching for one that satisfies the formulae in the theory.

However, we can greatly speed up this search if we instead impose structure on $D$.
Specifically, if we have predicates $P_1 \ldots P_N$, we take $D \subseteq \B^N$: each element of our domain is a boolean sequence
with a bit for each sub-expression $E$. 
The idea is that each element of $\B^N$ models a possible equivalence class of predicate truth values.
For $b \in D$, we want $b_i$ to be $\true$ when
$P_{E_i}(b)$ holds. This means that our maps $I_P$ are already fixed: 
$I_{P_{E_i}}(b) = b_i$ i.e. the $i$th bit of sequence $b$.

However, with this interpretation, $\B^N$ is too large to be our domain.
Suppose we have formulae $E_i$ and $E_j$ where $E_j = \lnot E_i$. 
Then there are sequences in $\B^n$ with both bits $i$ and $j$ set to $\true$.
To respect the consistency of our logic, we need $D$ to be a subset of $\B^N$
that eliminates such inconsistent elements.

Suppose that we have a function $\inDomain : \B^N -> \B $,
which determines whether a bit-sequence is in the domain of a potential model.
If $\Lvb{\bigwedge{\seq{L}}}$ contains the formula $\forall x_1 \in D \ldots \forall x_n \in D \ldotp \Phi[x_1 \ldots x_n]$,
for some $\Phi$, we can instead write:\\
${\forall b_1 \in \B^N \ldots \forall b_n \in \B^N \ldotp \inDomain(b_1) \land \ldots \land \inDomain(b_n) \implies \Phi[b_1 \ldots b_n]}$.\\
That is, our domain can only contain values that respect the semantics of set expressions.
Similarly, if $\Lvb{\bigwedge{\seq{L}}}$ contains $\exists x \ldotp \Phi[x]$, we can
write $\exists b \in \B^N \ldotp \inDomain(b) \land \Phi[b]$.
Since all functions in a model are implicitly closed over the domain, 
we also specify that $\forall \seq{b} \in (\B^n)^a \ldotp \seq{\inDomain(b)} \implies \inDomain(f_i^a(\seq{b}))$.
This ensures that our formulae over boolean sequences are equivalent to the original formulae.

This is enough to express $\Lvb{\bigwedge{\seq{L}}}$ as an SMT problem.
We assert the existence of $\inDomain : \B^N -> \B$ along with $f_i^a : (\B^N)^a \to \B^N$
for each function in our Herbrand universe.
We modify each formula 
in $\Lvb{\bigwedge{\seq{L}}}$ to constrain a boolean sequences variable $b_i \in \B^n$ in place of each variable $x_i \in D$ as described above.
We add $\inDomain$ qualifiers to  existentially and universally quantified formulae, and replace each $P_{E_i}(x_j)$ with the $i$th bit of $b_j$. 
We add a constraint asserting that each $f_i^a$ is closed over the values satisfying $\inDomain$.
The SMT solver searches for values for all existential variables, functions, and $\inDomain$
that satisfy this formula.

\subsection{Reducing the Search Space}

\begin{figure}[t]
  \begin{minipage}{0.45\textwidth}
  \begin{align*}
    &P_\top(b) & := &&& \true\\
    &P_{X}(b) & := &&& \text{bit for $X$ in $b$} \\
    &P_{f_i^a(E_1, \ldots E_a)}(b) & := &&& \text{bit for $f_i^a(E_1, \ldots E_a)$ in $b$}\\
    &P_{\lnot E_1}(b) & := &&& \lnot P_{E_1}(b)\\ 
  \end{align*}
\end{minipage}
\begin{minipage}{0.45\textwidth}
  \begin{align*}
    &P_\bot(b) & := &&& \false\\
    &P_{E_1 \cap E_2}(b) & := &&& P_{E_1}(b) \land P_{E_2}(b)\\
    &P_{E_1 \cup E_2}(b) & := &&& P_{E_1}(b) \lor P_{E_2}(b)\\
  \end{align*}
\end{minipage}
\caption{Recursive Definition of Predicates for the SMT Translation}
\label{fig:pred-defn}
\end{figure} 

While this translation corresponds nicely to the monadic translation,
it has more unknowns than are needed. Specifically, $\inDomain$ will always
reject boolean sequences that violate the constraints of each $\Eb{E_i}$.
For example, the bit for $P_{E_1 \cap E_2}$ in $b$ must always be exactly
$P_{E_1}(b) \land P_{E_2}(b)$. In fact, for each form except function applications
and set variables, the value of a bit for an expression can be recursively determined by values
of bits for its immediate subexpressions (\cref{fig:pred-defn}).
This means that our boolean sequences need only contain slots for expressions of the form
$X$ or $f_i^a(E_1, \ldots E_a)$, shrinking the problem's search space.

What's more, we now only need to include the constraints from $\Eb{}$ for 
expressions of the form $X$ or $f_i^a(E_1, \ldots E_a)$,
since the other constraints hold \textit{by definition} given our definitions of each $P_E$.
Similarly, our constraints restrict the freedom we have in choosing $f^a_i$.
Specifically, we know that $P_{f_i^a(E_1, \ldots, E_a)}(f_i^a(b_1, \ldots, b_a))$
should hold if and only if $P_{E_i}(b_i)$ holds for each $i \leq a$.
Similarly, we know that $P_{f_i^a(E_1, \ldots, E_a)}(g_j^{a'}(b_1, \ldots, b_{a'}))$
should always be $\false$ when $f\neq g$. So for each $f^a_i$, it suffices to find
a mapping from inputs $b_1, \ldots, b_a$ to the value of $P_X(f^a_i(b_1, \ldots, b_a))$
for each variable $X$.
This reduces the number of unknowns in the SMT problem.
 
\subsection{The Complete Translation}
\label{sec:complete-trans}

Given a conjunction of literals $\bigwedge \seq{L}$,
let $X_1, \ldots X_k, E_{k+1}, \ldots E_N$ be the sequence of variable and function-application
sub-expressions of $\seq{L}$. 
We define $P_E(b)$ for each sub-expression $E$ of $\seq{L}$ as in \cref{fig:pred-defn}.

As unknowns, we have:
\begin{easylist}[itemize]
  & a function $\inDomain : \B^N -> \B$;
  & for each negative literal $E_i \not \subseteq E'_i$, an existential variable $y_i \in \B^N$;
  & for each function $f^a_i$ and each variable $X \in \seq{L}$, 
  a function ${f^a_{iX} : (\B^N)^a -> \B}$, which takes $a$ sequences of $N$ bits, 
  and computes the value of the bit for $P_X$ in the result.
\end{easylist}

We define the following known functions:
\begin{easylist}[itemize]
  & 
  $f^a_{if_i^a(E_1, \ldots, E_a)} : (\B^N)^a -> \B$ for each $f^a_i$ and each sub-expression of the form $f_i^a(E_1, \ldots, E_a)$, where $f^a_{if_i^a(E_1, \ldots, E_a)}(b_1, \ldots, b_a) = P_{E_1}(b_1) \land \ldots\land  P_{E_a}(b_a)$;
     & $f^a_{ig_j^{a'}(E_1, \ldots, E_{a'})} : (\B^N)^a -> \B$ returning $\false$, for each $f^a_i$ and each sub-expression of the form $g_j^{a'}(E_1, \ldots, E_{a'})$
     where $f \neq g$;
     &  $f^a_{iSMT} : (\B^N)^a -> \B^N$ for each $f^a_i$, where 
      $f^a_{iSMT}(b_1, \ldots, b_a)$ is the sequence: \\
      ${f^a_{iX_1}(b_1, \ldots, b_a) \ldots f^a_{iX_k}(b_1, \ldots, b_a) f^a_{iE_{k+1}}(b_1, \ldots, b_a)\ldots f^a_{iE_{N}}(b_1, \ldots, b_a)}$
    \end{easylist}

  We assert that the following hold:
  \begin{easylist}[itemize]
    & for each negative constraint $E_i \not \subseteq E'_i$ with corresponding existential variable $y_i$,
   that  ${\inDomain(y_i) \land P_{E_i}(y_i) \land \lnot P_{E'_I}(y_i)} $ holds;
   & 
   ${\forall x \in \B^N \ldotp (\inDomain(x) \land P_{E_i}(x)) \implies P_{E'_i}(x)}$
   for each positive $E_i  \subseteq E'_i$;
   & $\forall x_{1} \ldots x_a \ldotp (\bigwedge_{j = 1 \ldots a } \inDomain(x_j))
   \!\implies\! \inDomain(f^a_{iSMT}(x_1, \ldots, x_a))$ for each function $f^a_i$ 
  \end{easylist}

A solution to these assertions exists iff the initial set constraint is satisfiable.

\subsection{Arbitrary Boolean Combinations}
\label{subsec:bool-comb}

Allowing arbitrary boolean combinations of set constraints enriches our pattern
match analysis and to allow us to use projections. To do this,
for each atom $E_i \subseteq E'_i$ in a constraint $C$, 
we introduce a boolean $\ell_i$, which the SMT solver guesses.
We modify our translation so that $\Lb{E_i \subseteq E'_i} = \ell_i \implies \forall x \ldotp (P_{E_i}(x) \implies P_{E'_i}(x))$
and $\Lb{E_i \not\subseteq E'_i} = \lnot\ell_i \implies (\exists y \ldotp P_{E_i}(y) \land \lnot P_{E'_i}(y))$.
So  $l_i$ is true iff $E_i \subseteq E'_i$.
Finally, we assert the formula that is $C$ where each occurrence of $E_i \subseteq E'_i$ is replaced by $\ell_i$
and $E_i \not\subseteq E'_i$ is replaced by $\lnot \ell_i$. 
Thus, we force our SMT solver to guess a literal assignment for each atomic set constraint,
and then determine if it can solve the conjunction of those literals.
When $\ell_i$ is false, then $\Lb{E_i \subseteq E'_i}$ will be vacuously true,
with the opposite holding for negative constraints.

\section{Evaluation and Discussion}
\label{sec:impl}

\begin{table*}[t!]   
  \begin{center}
  \caption{Compilation Time (ms) of Exhaustiveness versus Pattern Match Analysis}
  \label{tab:perf}
  \begin{tabular}{|l|cc|cc|cc|l|}
    \hline
    \textbf{Library} & \textbf{EX-TN} & \textbf{PMA-TN} & \textbf{EX-FP} & \textbf{PM-FP} & \textbf{EX-TP} & \textbf{PM-TP}   \\ 
    \hline 
    elm-graph & 50 & 168 & 45 & 178 & 44 & 173\\
elm-intdict & 42 & 115 & 38 & 5121* & 35 & 113\\ 
elm-interval & 40 & 69 & 39 & 1217* & 36 & 1261\\
    \hline
  \end{tabular}
\end{center}
\end{table*}

We implemented our translation~\citep{our-gh-solver} atop Z3 4.8.5 with \texttt{mbqi} and \texttt{UFBV}.
On an i7-3770 CPU 32GB RAM machine, we compared the running time of Elm's exhaustiveness check
with an implementation of our analysis~\citep{our-elm}.

In order to make the analysis practical, we implemented several optimizations
on top of our analysis. Trivially satisfiable constraints were removed,
and obvious simplifications were applied to set expressions.
When a match was exhaustive, its safety constraint was omitted,
and since non-safety constraints should be satisfiable, calls to Z3 were only made
for non-empty safety constraint lists.  
A union-find algorithm was used to combine variables constrained to be equal,
and intermediate variables were merged.
Since the constraint of an annotated scheme is copied at each instantiation,
these ensured that the size of type annotations did not explode. 
For simplicity, annotated types were not carried across module boundaries: imported functions
were assumed to accept any input and always have return annotation $\top$.
Similarly, a conservative approximation was used in place of the full projections
when determining pattern variables' annotations.

We ran our tests on the Elm graph\cite{elm-graph}, intdict\cite{elm-intdict}, 
and interval\cite{elm-interval} libraries. Each of these initially contained safe partial matches,
but were modified to  return dummy values in unreachable code when Elm 0.19 was released.
The results of the evaluation are given in \cref{tab:perf}.
Runs with the prefix \textbf{EX} used the exhaustiveness check of the original Elm compiler,
while those marked \textbf{PMA} used our pattern-match analysis.
We tested the compilers on three variants of each library, a true-negative (\textbf{-TN}) 
version in which all matches were exhaustive, a false-positive (\textbf{-FP}) version in which
a match was non-exhaustive but safe, and a true-positive (\textbf{-TP}) version in which a required branch was missing
and running the program would result in an error.
Cases marked with an asterisk (*) are those which were rejected by the Elm compiler,
but which our analysis marked as safe.
Notably, the elm-graph library relied on the invariant that a connected
component's depth-first search forest has exactly one element, which was too complex for our analysis to capture.        

Our analysis is slower than exhaustiveness checking in each case.
However, the pattern match analysis requires less than one second in the majority of cases,
and in the worst case requires only six seconds. The slowdown was most prominent in the false-positive cases
that our analysis marks as safe,
where Z3 was not able to quickly disprove the satisfiability of the constraints.
Conversely, in the \textbf{-TN} cases where Z3 was not called, our analysis cause very little slowdown.
Partial matches tend to occur rarely in code, so we feel this is acceptable performance
for a tool integrated into a compiler.

\textbf{Future Work:}
While our translation of set constraints to SMT attempts to minimize the search space,
we have not investigated further optimizations of the SMT problem.
The SMT solver was given
relatively small problems.
Few programs contain hundreds of constructors or pattern match cases.
Nevertheless, more can be done to reduce the time spent in the SMT solver for larger problems. 
Solvers like CVC4~\citep{BCD+11}
are highly configurable with regards to their strategies for solving quantification.
Fine tuning the configuration could decrease the times required to solve our problems without requiring a custom solver. 
Conversely, a solver specialized to quantified boolean arithmetic could yield faster results.

Likewise, type information could be used to speed up analysis.
While we have modeled patterns using the entire Herbrand space,
values of different data types reside in disjoint universes.
Accounting for this could help partition one problem with many variables
into several problems with few variables.

\textbf{Related Work - Set Constraints:}
The modern formulation of set constraints was established by \citet{113732}.
Several independent proofs of decidability for systems with negative constraints were given,
using a number-theoretic reduction \cite{AIKEN199530, systemsNeg}, tree automata \cite{treeAutomataSetNeg},
and monadic logic \cite{316078}.
Charatonik and Podelski established the decidability of positive and negative constraints with projection \cite{setConstraintsProjection}.
The first tool aimed at a general, practical solver for set constraints
was BANE \cite{10.1007/BFb0055513}, which used a system of rewrite
rules to solve a restricted form of set constraints \cite{Aiken:1993:TIC:165180.165188}.
Banshee improved BANE's performance with code generation
and incremental analysis \cite{10.1007/11547662_16}.
Neither of these implementations allow for  negative constraints or unrestricted projections.
Several survey papers give a more in-depth overview
of set constraint history and research~\citep{setConstraintPearl,AIKEN199979, 10.1007/3-540-58601-6_107}.
 
\textbf{Related Work -  Pattern Match Analysis:}
Several pattern match analyses have been presented in previous work.
\citet{Koot:Thesis:2012} presents a higher-order pattern match analysis as a type-and-effect system,
using a presentation similar to ours.
This work was extended by \citet{Koot:2015:TEA:2678015.2682542}, who present an analysis based on
higher-order polymorphism. This improves the precision of the analysis,
but suffers from the same problems as our regarding polymorphic recursion.
All of these efforts use restricted versions of set constraints, and do not allow for
unrestricted projection, negation, and boolean combinations of constraints.

Previous versions of type inference for pattern matching have utilized
\textit{conditional
  constraints}~\citep{Aiken:1994:STC:174675.177847,Pottier:2000:VCT:763845.763849,
  10.1007/978-3-319-12736-1_6},  similar to our path constraints.
\citet{Castagna:2015:PFS:2676726.2676991} describe a similar system, albeit more focused on type-case than
pattern matching.
Catch~\citep{10.1145/1543134.1411293} uses a similar system
of \textit{entailment}, with a restricted constraint language to ensure finiteness.
These systems are similar in expressive power to the constraints that we used in our final implementation,
but our underlying constraint logic is more powerful.
There are restrictions on where unions and intersections can appear in conditional constraints~\citep{Aiken:1994:STC:174675.177847},
and there is not full support for projections or negative constraints.
In particular, negative constraints allow for analyses to specify that a function's input
set must not be empty, so that the type error can point to the function definition rather than the call-site,
avoiding the ``lazy'' inference described by \citet{Pottier:2000:VCT:763845.763849}.   
While these have not been integrated into our implementation, our constraint logic makes it easy to incorporate these
and other future improvements. 

Another related line of work is \textit{datasort refinements}.
\citep{Freeman:1991:RTM:113445.113468, Dunfield:2003:TAI:1754809.1754827, Dunfield:2004:TT:964001.964025,Dunfield:2007:RTS:1292597.1292602}. 
As with our work, the goal of datasort refinements is to allow partial pattern matches
while eliminating runtime failures. This is achieved by introducing \textit{refinements}
of each algebraic data type corresponding to its constructors, possibly with unions or
intersections.
Datasort refinements are presented as a type system,
not as a standalone analysis, so their handling of polymorphism and recursive types 
is more precise than ours. However,
checking programs with refined types requires at least some annotation from the programmer, where our analysis
can check programs without requiring additional programmer input.

\textbf{Conclusion:}
Unrestricted set constraints previously were used only in theory.
With our translation, they can be solved in practice.
SMT solvers are a key tool in modern verification, and they can now be used
to solve set constraints.
We have shown that even $\NEXPTIME$-completeness is not a complete
barrier to the use of set constraints in practical verification.

\bibliographystyle{splncsnat}
\bibliography{myrefs}

\begin{thebibliography}{38}
\providecommand{\natexlab}[1]{#1}
\providecommand{\url}[1]{\texttt{#1}}
\providecommand{\urlprefix}{}

\bibitem[{Ackermann(1954)}]{Ackermann}
Ackermann, W.: Solvable cases of the decision problem.
\newblock Studies in logic and the foundations of mathematics, North-Holland
  Pub. Co. (1954)

\bibitem[{Aiken et~al.(1995)Aiken, Kozen, and Wimmers}]{AIKEN199530}
Aiken, A., Kozen, D., Wimmers, E.: Decidability of systems of set constraints
  with negative constraints.
\newblock Information and Computation 122(1), 30 -- 44 (1995),
  \urlprefix\url{http://www.sciencedirect.com/science/article/pii/S089054018571139X}

\bibitem[{Aiken(1999)}]{AIKEN199979}
Aiken, A.: Introduction to set constraint-based program analysis.
\newblock Science of Computer Programming 35(2), 79 -- 111 (1999),
  \urlprefix\url{http://www.sciencedirect.com/science/article/pii/S0167642399000076}

\bibitem[{Aiken et~al.(1998)Aiken, F{\"a}hndrich, Foster, and
  Su}]{10.1007/BFb0055513}
Aiken, A., F{\"a}hndrich, M., Foster, J.S., Su, Z.: A toolkit for constructing
  type- and constraint-based program analyses.
\newblock In: Leroy, X., Ohori, A. (eds.) Types in Compilation. pp. 78--96.
  Springer Berlin Heidelberg, Berlin, Heidelberg (1998)

\bibitem[{Aiken and Wimmers(1993)}]{Aiken:1993:TIC:165180.165188}
Aiken, A., Wimmers, E.L.: Type inclusion constraints and type inference.
\newblock In: Proceedings of the Conference on Functional Programming Languages
  and Computer Architecture. pp. 31--41. FPCA '93, ACM, New York, NY, USA
  (1993), \urlprefix\url{http://doi.acm.org/10.1145/165180.165188}

\bibitem[{Aiken et~al.(1994)Aiken, Wimmers, and
  Lakshman}]{Aiken:1994:STC:174675.177847}
Aiken, A., Wimmers, E.L., Lakshman, T.K.: Soft typing with conditional types.
\newblock In: Proceedings of the 21st ACM SIGPLAN-SIGACT Symposium on
  Principles of Programming Languages. pp. 163--173. POPL '94, ACM, New York,
  NY, USA (1994), \urlprefix\url{http://doi.acm.org/10.1145/174675.177847}

\bibitem[{{Bachmair} et~al.(1993){Bachmair}, {Ganzinger}, and
  {Waldmann}}]{287598}
{Bachmair}, L., {Ganzinger}, H., {Waldmann}, U.: Set constraints are the
  monadic class.
\newblock In: [1993] Proceedings Eighth Annual IEEE Symposium on Logic in
  Computer Science. pp. 75--83 (June 1993)

\bibitem[{Barrett et~al.(2011)Barrett, Conway, Deters, Hadarean,
  Jovanovi{'{c}}, King, Reynolds, and Tinelli}]{BCD+11}
Barrett, C., Conway, C.L., Deters, M., Hadarean, L., Jovanovi{'{c}}, D., King,
  T., Reynolds, A., Tinelli, C.: {CVC4}.
\newblock In: Gopalakrishnan, G., Qadeer, S. (eds.) Proceedings of the 23rd
  International Conference on Computer Aided Verification (CAV '11). Lecture
  Notes in Computer Science, vol. 6806, pp. 171--177. Springer (Jul 2011),
  \urlprefix\url{http://www.cs.stanford.edu/~barrett/pubs/BCD+11.pdf}

\bibitem[{Bell(2019)}]{elm-interval}
Bell, R.K.: elm-interval.
\newblock \url{https://github.com/r-k-b/elm-interval/} (2019), commit a7f5f8a

\bibitem[{Castagna et~al.(2015)Castagna, Nguyen, Xu, and
  Abate}]{Castagna:2015:PFS:2676726.2676991}
Castagna, G., Nguyen, K., Xu, Z., Abate, P.: Polymorphic functions with
  set-theoretic types: Part 2: Local type inference and type reconstruction.
\newblock In: Proceedings of the 42Nd Annual ACM SIGPLAN-SIGACT Symposium on
  Principles of Programming Languages. pp. 289--302. POPL '15, ACM, New York,
  NY, USA (2015), \urlprefix\url{http://doi.acm.org/10.1145/2676726.2676991}

\bibitem[{{Charatonik} and {Pacholski}(1994)}]{316078}
{Charatonik}, W., {Pacholski}, L.: Negative set constraints with equality.
\newblock In: Proceedings Ninth Annual IEEE Symposium on Logic in Computer
  Science. pp. 128--136 (July 1994)

\bibitem[{Czaplicki(2019)}]{elmGuide}
Czaplicki, E.: Introduction to {Elm} (2019),
  \urlprefix\url{http://guide.elm-lang.org/}

\bibitem[{Damas and Milner(1982)}]{Damas:1982:PTF:582153.582176}
Damas, L., Milner, R.: Principal type-schemes for functional programs.
\newblock In: Proceedings of the 9th ACM SIGPLAN-SIGACT Symposium on Principles
  of Programming Languages. pp. 207--212. POPL '82, ACM, New York, NY, USA
  (1982), \urlprefix\url{http://doi.acm.org/10.1145/582153.582176}

\bibitem[{Dunfield(2007)}]{Dunfield:2007:RTS:1292597.1292602}
Dunfield, J.: Refined typechecking with {Stardust}.
\newblock In: Proceedings of the 2007 Workshop on Programming Languages Meets
  Program Verification. pp. 21--32. PLPV '07, ACM, New York, NY, USA (2007),
  \urlprefix\url{http://doi.acm.org/10.1145/1292597.1292602}

\bibitem[{Dunfield and Pfenning(2003)}]{Dunfield:2003:TAI:1754809.1754827}
Dunfield, J., Pfenning, F.: Type assignment for intersections and unions in
  call-by-value languages.
\newblock In: Proceedings of the 6th International Conference on Foundations of
  Software Science and Computation Structures and Joint European Conference on
  Theory and Practice of Software. pp. 250--266. FOSSACS'03/ETAPS'03,
  Springer-Verlag, Berlin, Heidelberg (2003),
  \urlprefix\url{http://dl.acm.org/citation.cfm?id=1754809.1754827}

\bibitem[{Dunfield and Pfenning(2004)}]{Dunfield:2004:TT:964001.964025}
Dunfield, J., Pfenning, F.: Tridirectional typechecking.
\newblock In: Proceedings of the 31st ACM SIGPLAN-SIGACT Symposium on
  Principles of Programming Languages. pp. 281--292. POPL '04, ACM, New York,
  NY, USA (2004), \urlprefix\url{http://doi.acm.org/10.1145/964001.964025}

\bibitem[{Eremondi(2019{\natexlab{a}})}]{our-elm}
Eremondi, J.: Forked elm-compiler.
\newblock \url{https://github.com/JoeyEremondi/elm-compiler-patmatch-smt}
  (2019{\natexlab{a}}), commit 9581aaf

\bibitem[{Eremondi(2019{\natexlab{b}})}]{our-gh-solver}
Eremondi, J.: Setconstraintssmt.
\newblock \url{https://github.com/JoeyEremondi/SetConstraintsSMT}
  (2019{\natexlab{b}}), commit 03bb754

\bibitem[{Freeman and Pfenning(1991)}]{Freeman:1991:RTM:113445.113468}
Freeman, T., Pfenning, F.: Refinement types for {ML}.
\newblock In: Proceedings of the ACM SIGPLAN 1991 Conference on Programming
  Language Design and Implementation. pp. 268--277. PLDI '91, ACM, New York,
  NY, USA (1991), \urlprefix\url{http://doi.acm.org/10.1145/113445.113468}

\bibitem[{Gilleron et~al.(1993)Gilleron, Tison, and
  Tommasi}]{treeAutomataSetNeg}
Gilleron, R., Tison, S., Tommasi, M.: Solving systems of set constraints with
  negated subset relationships.
\newblock In: Proceedings of 1993 IEEE 34th Annual Foundations of Computer
  Science. pp. 372--380 (Nov 1993)

\bibitem[{Graf(2019{\natexlab{a}})}]{elm-graph}
Graf, S.: elm-graph.
\newblock \url{https://github.com/elm-community/graph/} (2019{\natexlab{a}}),
  commit 3672c75

\bibitem[{Graf(2019{\natexlab{b}})}]{elm-intdict}
Graf, S.: elm-intdict.
\newblock \url{https://github.com/elm-community/intdict} (2019{\natexlab{b}}),
  commit bf2105d

\bibitem[{{Heintze} and {Jaffar}(1990)}]{113732}
{Heintze}, N., {Jaffar}, J.: A decision procedure for a class of set
  constraints.
\newblock In: [1990] Proceedings. Fifth Annual IEEE Symposium on Logic in
  Computer Science. pp. 42--51 (June 1990)

\bibitem[{Heintze and Jaffar(1994)}]{10.1007/3-540-58601-6_107}
Heintze, N., Jaffar, J.: Set constraints and set-based analysis.
\newblock In: Borning, A. (ed.) Principles and Practice of Constraint
  Programming. pp. 281--298. Springer Berlin Heidelberg, Berlin, Heidelberg
  (1994)

\bibitem[{Klabnik and Nichols(2018)}]{klabnik2018rust}
Klabnik, S., Nichols, C.: The Rust Programming Language.
\newblock No Starch Press (2018)

\bibitem[{Kodumal and Aiken(2005)}]{10.1007/11547662_16}
Kodumal, J., Aiken, A.: Banshee: A scalable constraint-based analysis toolkit.
\newblock In: Hankin, C., Siveroni, I. (eds.) Static Analysis. pp. 218--234.
  Springer Berlin Heidelberg, Berlin, Heidelberg (2005)

\bibitem[{Koot(2012)}]{Koot:Thesis:2012}
Koot, R.: {Higher-Order Pattern Match Analysis}.
\newblock Master's thesis, Universiteit Utrecht, the Netherlands (2012)

\bibitem[{Koot and Hage(2015)}]{Koot:2015:TEA:2678015.2682542}
Koot, R., Hage, J.: Type-based exception analysis for non-strict higher-order
  functional languages with imprecise exception semantics.
\newblock In: Proceedings of the 2015 Workshop on Partial Evaluation and
  Program Manipulation. pp. 127--138. PEPM '15, ACM, New York, NY, USA (2015),
  \urlprefix\url{http://doi.acm.org/10.1145/2678015.2682542}

\bibitem[{L{\"o}wenheim(1915)}]{Lowenheim1915}
L{\"o}wenheim, L.: {\"U}ber m{\"o}glichkeiten im relativkalk{\"u}l.
\newblock Mathematische Annalen 76(4), 447--470 (Dec 1915),
  \urlprefix\url{https://doi.org/10.1007/BF01458217}

\bibitem[{Mitchell and Runciman(2008)}]{10.1145/1543134.1411293}
Mitchell, N., Runciman, C.: Not all patterns, but enough: An automatic verifier
  for partial but sufficient pattern matching.
\newblock SIGPLAN Not. 44(2), 49–60 (Sep 2008),
  \urlprefix\url{https://doi.org/10.1145/1543134.1411293}

\bibitem[{de~Moura and Bj{\o}rner(2008)}]{10.1007/978-3-540-78800-3_24}
de~Moura, L., Bj{\o}rner, N.: Z3: An efficient {SMT} solver.
\newblock In: Ramakrishnan, C.R., Rehof, J. (eds.) Tools and Algorithms for the
  Construction and Analysis of Systems. pp. 337--340. Springer Berlin
  Heidelberg, Berlin, Heidelberg (2008)

\bibitem[{Nielson and Nielson(1999)}]{Nielson:1999:TES:646005.673740}
Nielson, F., Nielson, H.R.: Type and effect systems.
\newblock In: Correct System Design, Recent Insight and Advances, (to Hans
  Langmaack on the Occasion of His Retirement from His Professorship at the
  University of Kiel). pp. 114--136. Springer-Verlag, Berlin, Heidelberg
  (1999), \urlprefix\url{http://dl.acm.org/citation.cfm?id=646005.673740}

\bibitem[{Pacholski and Podelski(1997)}]{setConstraintPearl}
Pacholski, L., Podelski, A.: Set constraints: A pearl in research on
  constraints, pp. 549--561.
\newblock Springer Berlin Heidelberg, Berlin, Heidelberg (1997),
  \urlprefix\url{https://doi.org/10.1007/BFb0017466}

\bibitem[{Pacholski(2010)}]{setConstraintsProjection}
Pacholski, W.C.L.: Set constraints with projections.
\newblock J. ACM 57(4), 23:1--23:37 (May 2010),
  \urlprefix\url{http://doi.acm.org/10.1145/1734213.1734217}

\bibitem[{Palmer et~al.(2014)Palmer, Menon, Rozenshteyn, and
  Smith}]{10.1007/978-3-319-12736-1_6}
Palmer, Z., Menon, P.H., Rozenshteyn, A., Smith, S.: Types for flexible
  objects.
\newblock In: Garrigue, J. (ed.) Programming Languages and Systems. pp.
  99--119. Springer International Publishing, Cham (2014)

\bibitem[{Pottier(2000)}]{Pottier:2000:VCT:763845.763849}
Pottier, F.: A versatile constraint-based type inference system.
\newblock Nordic J. of Computing 7(4), 312--347 (Dec 2000),
  \urlprefix\url{http://dl.acm.org/citation.cfm?id=763845.763849}

\bibitem[{Stefansson(1994)}]{systemsNeg}
Stefansson, K.: Systems of set constraints with negative constraints are
  {NEXPTIME}-complete.
\newblock In: Proceedings Ninth Annual IEEE Symposium on Logic in Computer
  Science. pp. 137--141 (Jul 1994)

\bibitem[{Tseitin(1983)}]{Tseitin1983}
Tseitin, G.S.: On the Complexity of Derivation in Propositional Calculus, pp.
  466--483.
\newblock Springer Berlin Heidelberg, Berlin, Heidelberg (1983),
  \urlprefix\url{https://doi.org/10.1007/978-3-642-81955-1_28}

\end{thebibliography}

\end{document}